\documentclass[12pt]{article}
\usepackage{fullpage}
\usepackage{setspace}
\usepackage{comment}
\usepackage[pdftex]{graphicx}
\usepackage{rotating}
\usepackage{amsmath}
\usepackage{amsthm}
\usepackage{amssymb}
\usepackage{graphicx}
\usepackage{fullpage}
\usepackage{setspace}
\usepackage{natbib}
\usepackage{color}

\newcommand{\bHatm}{\hat{\beta}_{m}}

\newcommand{\bHat}{\hat{\beta}_n}

\newcommand{\rp}{\mathbb{R}^p}

\newcommand{\T}{\intercal}

\theoremstyle{definition}

\begin{document}

\title{An imputation method for estimating the learning curve in
  classification problems}

\author{Eric B. Laber\\Department of Statistics\\North Carolina State
University
  \and Kerby Shedden\\Department of Statistics\\University of Michigan
  \and Yang Yang\\ Yahoo! Research Labs}

\maketitle

\pagebreak

\begin{abstract}
The learning curve expresses the error rate of a predictive modeling
procedure as a function of the sample size of the training dataset. It
typically is a decreasing, convex function with a positive limiting
value. An estimate of the learning curve can be used to assess whether
a modeling procedure should be expected to become substantially more
accurate if additional training data become available. This article
proposes a new procedure for estimating learning curves using
imputation. We focus on classification, although the idea is
applicable to other predictive modeling settings. Simulation studies
indicate that the learning curve can be estimated with useful accuracy
for a roughly four-fold increase in the size of the training set
relative to the available data, and that the proposed imputation
approach outperforms an alternative estimation approach based on
parameterizing the learning curve. We illustrate the method with an
application that predicts the risk of disease progression for people
with chronic lymphocytic leukemia.
\end{abstract}

\pagebreak

\section{Introduction}

Predictive models describe the relationship between an outcome and a
set of predictor variables, and are widely used in areas ranging from
personalized medicine to computational advertising. For example, in
personalized medicine the aim may be to predict which patients with a
particular disease are likely to respond favorably to a treatment
based on information contained in a set of pre-treatment biomarkers
(\cite{insel2009translating}). Predictive models are developed using a
training data set, and their ``generalization performance'' is
typically assessed with respect to a test set that is independent of
the training set. Minimizing some measure of the generalization error
rate is usually the first priority in predictive modeling, although
issues such as model simplicity, interpretability, and ease of
implementation may also be important.

One common type of predictive modeling is binary classification, in
which the outcome $Y$ takes on one of two possible values. In
classification problems, the expected misclassification rate is a
natural measure of generalization performance. Suppose we observe a
training set of $n$ feature-label pairs $\mathcal{D}_{n} = \lbrace
(X_i, Y_i) \rbrace_{i=1}^n$ (capital letters such as $X$ and $Y$ 
denote random variables and lower case letters such as $x$ and $y$
denote instances of these random variables throughout).  Using the
training data $\mathcal{D}_{n}$, we can construct a classifier
$\hat{c}_{n}$, say using logistic regression. The goal is to use the
classifier $\hat{c}_{n}$ to accurately predict the labels $Y$ from the
observed features $X$ of unlabeled cases. The expected
misclassification rate $\tau(n)$ of the classifier $\hat{c}_{n}$ is
the expected proportion of incorrectly labeled features $X$ averaged
over both the feature-label distribution of test cases and the
distribution of $\mathcal{D}_{n}$; that is,

\begin{equation}\label{expectedErrorRate}
\tau(n) \triangleq \mathbb{E}\lbrace Y \ne \hat{c}_{n}(X)\rbrace =
E_{\mathcal{D}_n} E\lbrace{Y \ne \hat{c}_{n}(X)|\mathcal{D}_{n}
  \rbrace},
\end{equation}

\noindent where $\mathbb{E}$ denotes expectation taken with respect to
both $(X,Y)$ and the training data $\mathcal{D}_{n}$. Our primary
focus here is to estimate the function $\tau(n)$, which has been
termed the ``learning curve'' (\cite{amari1992}, \cite{haussler1996},
\cite{trevor2009elements} page 243).

Knowledge of the learning curve can contribute to both study design
and interpretation of predictive modeling results. Study design
questions arise naturally if the training data are acquired in two or
more stages, or were obtained from a pilot study with a modest sample
size. Such a study may produce encouraging evidence that a useful
predictive relationship exists, but one naturally expects that a
predictive model obtained from a small training set will not perform
as well as one obtained using a larger training set. If we learn that
the expected generalization performance of a rule obtained by
following the study design employed in our pilot study can be
substantially improved by using a larger training set, we would be
encouraged to conduct a larger study using the same features and
predictive modeling approach.

Learning curves can also contribute to the interpretation of
predictive modeling results. In many settings in which predictive
modeling is applied, the variables naturally fall into domains. For
example, many early genomic studies investigating risk prediction for
cancer outcomes used the expression levels of genes associated with
cell growth, division, and proliferation to inform the predictions.
When considering the performance of these early studies, which often
had modest sample sizes, it was natural to ask whether their
performance could best be improved by using larger training sets, or
by considering additional classes of genes such as those involved in
resistance to chemotherapeutic agents or inhibition of the immune
response. A similar question arises when considering the integration
of data from different domains that may influence disease outcomes,
such as environmental influences, measured metabolite levels, and
inherited genetic factors.

The rest of the paper is organized as follows. Section \ref{lcreview}
reviews some previous work on learning curves.  Section \ref{lcest}
describes three approaches to learning curve estimation, including one
existing approach and two new approaches. Section \ref{simstudy}
compares the performances of these approaches using several simulated
examples. Section \ref{cll} illustrates the imputation approach using
a data set in which the goal is to predict the risk that a patient
with chronic lymphocytic leukemia (CLL) will experience a poor
outcome. Section \ref{discussion} provides some concluding remarks.

\section{Learning curves}\label{lcreview}

Learning curves have been an object of interest for several decades.
Bounds on the learning curve follow from the work of Vapnik and
Chervonenkis (\cite{vapnik1971},\cite{vapnik1982}).  These bounds have
the power law form $a + b/m^\alpha$, where $\alpha=1/2$ holds in the
most realistic settings.  The bounds are tight if nothing is known
about the distribution of $X$ and one takes a worst-case perspective.
If information about the distribution of $X$ is available or can be
estimated from observed $X$ values, tighter bounds can be obtained
(e.g.\ \cite{haussler1996}).

The problem we consider here is to estimate the learning curve rather
than to bound it.  Thus we consider the setting in which data on
$(X,Y)$ are available, on which the estimation can be based, and we
focus on traditional criteria for statistical estimation such as bias
and variance, rather than on obtaining bounds.  To succeed at this
estimation, we must capture the general form of the function
(e.g.\ the order at which the function $\tau$ changes with $n$), but
also the relevant constants and lower order terms.

Our focus here is on binary classification, but for comparison we
briefly consider the setting of linear regression using least-squares
methods.  In this case, an expression for the learning curve $\tau(n)$
can be derived explicitly. The generalization performance in this case
is naturally assessed using the mean-square prediction error (MSPE)
$E[(Y-\hat{Y})^2]$ for a prediction $\hat{Y} = \hat{Y}(x)$ of the
unobserved $Y$ given its feature vector $X=x$. For training sets of
sample size $n$, the expected MSPE is $\sigma^2(1 + {\rm tr}(CM)/n)$,
where $M=E[(X^\prime X/n)^{-1}]$ for the training design matrix $X$
and $C=E[x^*x^{*\prime}]$ for the test set covariate vector $x^*$. The
reduction in MSPE due to the use of a larger training set is reflected
in the term $\sigma^2c/n$, where $c = {\rm tr}(CM)$ captures both the
complexity of the model and the similarity of the training and testing
distributions of covariate vectors.  We note that on the more natural
scale ${\rm RMSPE} = [{\rm MSPE}]^{1/2}$, this learning curve would
have the form $a + b/n^{1/2}$.

\section{Approaches to learning curve estimation}\label{lcest}

In this section, we describe three approaches to learning curve
estimation. The first approach follows a proposal of
\cite{mukherjee2003estimating}. The second and third approaches are
new, to our knowledge.

\subsection{Estimating the learning curve via subsampling and extrapolation}

In 2003, Mukherjee et al.\ described an approach to learning curve
estimation based on parameterizing the learning curve. We are not
aware that a name has been given to this method, and therefore we
termed it ``SUBEX'' for ``subsampling and extrapolation.'' The method
parameterizes the learning curve as an inverse power law of the form
$\tau(m) = a + bm^{-\alpha}$.  As noted above, this expression is
exact in the case of linear least squares regression, but may be
inexact in the case of classification using logistic regression.  The
unknown parameters in this expression are $a\in\mathbb{R}$ and
$b,\,\alpha \ge 0$. This parametric form is fit by first using
cross-validation on subsamples of the data of various sizes $m^\prime
< m$ to obtain direct estimates of $\tau(m^\prime)$. Specifically, for
a given $m^\prime < m$, we can subsample $B$ subsets of the training
data of size $m^\prime$, fit the classification model to each subset,
and use the complementary $m-m^\prime$ samples to unbiasedly estimate
the error rate. These $B$ error rate estimates can be averaged to
estimate $\tau(m^\prime)$. The parametric form for $\tau(m)$ is then
fit to these values to estimate $a, b$, and $\alpha$ using some form
of nonlinear regression.  For example, nonlinear least squares would
estimate $a$, $b$ and $\alpha$ by minimizing

$$ \sum_k (\hat{\tau}(m_k) - a - bm_k^{-\alpha})^2,$$

\noindent where the $m_k\le m$ are a set of sample sizes on which the
error rate is directly estimated.

As shown in Section \ref{lcest}, the SUBEX estimator can be positively
biased, conveying an overly optimistic assessment of the
generalization performance.  One reason for this optimism is the
asymmetry in the constraints placed on $\tau$.  The curve is
constrained to be non-increasing, which is quite natural, but owing to
the high variance in cross-validation estimators, the constraint is
active on a non-negligible proportion of modest-sized training sets
under simple generative models. By contrast, aside from being
constrained to be non-negative, the curve is unrestricted in how
rapidly it can decrease. Thus, when averaged over training sets, a
negative bias in the learning curve results.

\subsection{Estimating the learning curve via imputation and interpolation}

The second approach we consider for estimating the learning curve uses
data imputation and interpolation, hence this approach is termed
``IMPINT.'' In this approach, one first estimates the joint
distribution of the feature-label pair $(X, Y)$. This estimation is
generally performed by separately estimating the feature distribution
$p_{X}$, and the conditional distribution $p_{Y|X}$ of the label $Y$
given the feature $X$. After estimating the joint distribution, one
can synthesize data sets of any size. The ability to simulate such
data sets allows direct estimation of any point on the learning
curve. Specifically, one can generate an arbitrary number of training
sets of a given size $m$ from $\hat{p}_X\hat{p}_{Y|X}$, build a
classifier on each one, and then average the generalization
performance over newly drawn feature-label pairs from
$\hat{p}_X\hat{p}_{Y|X}$. The complete learning curve can be obtained
via interpolation among the learning curve points that are directly
estimated.

We now describe in detail the IMPINT procedure based on logistic
regression. Suppose we observe a training set $\mathcal{D}_{n}$ of
feature-label pairs $\lbrace (X_i, Y_i) \rbrace_{i=1}^{n}$ drawn
$i.i.d.$ from an unknown distribution with density $p_{X,Y}(x,y)$. The
features $X$ take values in $\rp$, and the binary labels $Y$ are coded
to take values in $\lbrace 0, 1\rbrace$. Define $\pi(x;\beta)
\triangleq \mathrm{logit}(x^{\T}\beta) = 1/(1+e^{-x^{\T}\beta})$. The
distribution $p_{X,Y}(x,y)$ factors into the product
$p_{Y|X}(y|x)p_{X}(x)$.  Under the logistic regression model the
conditional distribution of $Y|X$ has the form $p_{Y|X}(y|x) =
\pi(x;\beta^*)^{y}(1-\pi(x;\beta^*))^{1-y}$, with $\beta^* \in \rp$
denoting the unknown true parameter value. The marginal distribution
of the label $X$, denoted by $p_{X}(x)$, is left unspecified for now.

The conditional error rate for a particular training set
$\mathcal{D}_{m}$ is

\begin{equation}\label{innerExpectation}
R(\mathcal{D}_{m};\beta^*) \triangleq \int\left[ \pi(x;\beta^*)1\lbrace
  x^{\T}\hat{\beta}_{m} \le \kappa\rbrace + (1-\pi(x;\beta^*))1\lbrace
  x^{\T}\hat{\beta}_{m} > \kappa\rbrace \right]p_{X}(x)dx,
\end{equation}

\noindent where $\hat{\beta}_{m} = \hat{\beta}_{m}(\mathcal{D}_{m})$
is the maximum likelihood estimator of $\beta^*$ and the
classification rule is given by $\hat{c}_{m}(x) = 1\lbrace
x^{\T}\hat{\beta}_{m} \ge \kappa\rbrace$ for some threshold $\kappa
\in \mathbb{R}$. Using (\ref{expectedErrorRate}), we can express the
learning curve as

\begin{equation}\label{densityExpectedErrorRate}
\tau(m) = \mathbb{E} R(\mathcal{D}_{m}, \beta^*)= \int
R(\mathcal{D}_{m},
\beta^*)\prod_{i=1}^{m}p_{X}(x_i)p_{Y|X}(y_i)dx_idy_i.
\end{equation}

If $\beta^*$ and $p_{X}(x)$ were both known, one could compute
$\tau(m)$ for each $m$ using (\ref{densityExpectedErrorRate}), for
example, using Monte Carlo methods to approximate the
$2m+1$-dimensional integral with arbitrary accuracy. More
specifically, using $\beta^*$ and $p_{X}(x)$ one could generate $B$
training sets $\mathcal{D}_{m}^{(1)}, \mathcal{D}_{m}^{(2)}, \ldots,
\mathcal{D}_{m}^{(B)}$, each of size $m$. Fitting a logistic
regression model on the $b^{\rm th}$ training set yields the estimator
$\hat{\beta}_{m}^{(b)}$ of $\beta^*$. Furthermore, one could use
$\beta^*$ and $p_{X}(x)$ to generate a large test set $\mathcal{D}_*$.
For sufficiently large $B$ and $N$, it follows that
 
\begin{align}
   \tau(m) &\approx \frac{1}{BN}\sum_{b=1}^{B} \sum_{(X,Y)\in
     \mathcal{D}_*}\left[ \pi(X;\beta^*)\cdot 1\lbrace
     X^{\T}\bHatm^{(b)} \le \kappa\rbrace + (1-\pi(X;\beta^*))\cdot
     1 \lbrace X^{\T}\bHatm^{(b)} > \kappa\rbrace \right]
   \nonumber \\ &\approx
   \frac{1}{BN}\sum_{b=1}^{B}\sum_{(X,Y)\in\mathcal{D}_*}\left[ Y\cdot
     1\lbrace X^{\T}\bHatm^{(b)} \le \kappa\rbrace + (1-Y)\cdot
     1\lbrace X^{\T}\bHatm^{(b)} > \kappa \rbrace \right].\label{mc}
 \end{align}
 
The IMPINT estimate of $\tau(m)$, which we denote by
$\hat{\tau}_{II}(m)$, is formed by applying the approximation given in
(\ref{mc}) over imputed training and testing sets generated using an
imputation model fit to the complete observed training data
$\mathcal{D}_{n}$. More specifically, let $\hat{p}_{X}(x)$ denote an
estimator of $p_{X}(x)$ and $\bHat$ denote the usual maximum
likelihood estimator of $\beta^*$. Note that labeled data is not
needed to estimate $p_{X}(x)$, hence if additional unlabeled features
are available they can be used to improve the estimation of
$p_{X}(x)$. The estimators $\hat{p}_{X}(x)$ and $\bHat$, substituted
for $p_{X}(x)$ and $\beta^*$ respectively, can be used to impute $B$
training sets $\hat{\mathcal{D}}_{m}^{(1)},
\hat{\mathcal{D}}_{m}^{(2)}, \ldots, \hat{\mathcal{D}}_{m}^{(B)}$,
each of size $m$, and to sample a large synthetic test set
$\hat{\mathcal{D}}_*$. Using (\ref{mc}), the IMPINT estimator is given
by

\begin{equation}\label{piEstimatorM}
   \hat{\tau}_{II}(m) \triangleq \frac{1}{BN}\sum_{b=1}^B
   \sum_{(X,Y)\in \hat{\mathcal{D}}_*}\left[ Y\cdot 1\lbrace
     X^{\T}\bHatm^{(b)} \le \kappa\rbrace + (1-Y)\cdot 1\lbrace
     X^{\T}\bHatm^{(b)} > \kappa\rbrace \right],
\end{equation}

\noindent where $\bHatm^{(b)} =
\bHatm^{(b)}(\hat{\mathcal{D}}_{m}^{(b)})$ denotes the maximum
likelihood estimator based on the $b^{\rm th}$ imputed training set
$\hat{\mathcal{D}}_{m}^{(b)}$.

The model $\hat{p}_X$ for $p_X$ can be obtained using any appropriate
modeling approach for multivariate data. Some possible approaches are
demonstrated in the simulation studies and real data analysis below.
To avoid model mis-specification, it is tempting to simply use the
empirical distribution function of $X$ in place of $p_{X}$.  However,
in our experience, this approach does not work well with continuous
covariates. In particular, the estimated learning curve tends be
substantially biased downward. This may occur because points in the
training set have positive mass in the testing population. Thus, the
model is not relied upon to interpolate probabilities between observed
$X$ points; see \cite{efron1983estimating} for a discussion of the
role played by the distance between training and testing sets in
classification.

We found that the number of imputed data sets required to ensure that
the IMPINT estimate is smoothly non-increasing can be relatively
large.  As a practical matter, it is more computationally efficient to
use a smaller value of $B$ (e.g.\ $B\approx 500-1000$), and then feed
the resulting estimate through a monotone smoother (e.g., Friedman and
Tibshirani 1984; alternatively a parametric model could be fit as in
Mukherjee et al. 2003). We found that the use of a monotone smoother
reduces variance without introducing detectable additional bias.

\subsection{Bias reduction for learning curve estimates}

We found that a simple bias reduction substantially improves the
performance of the learning estimates obtained from imputed data.
This leads to a modified IMPINT approach, which we call BRIE for
``Bias Reduced Imputation Estimator.'' To motivate this approach,
consider what happens when we estimate $p_{Y|X}$ using the
best-fitting regression model (e.g., a fitted logistic regression
model). This will overstate the strength of the relationship between
$X$ and $Y$, particularly when the true relationship is weak (e.g., if
$Y$ and $X$ are independent, then $\hat{p}_{Y|X}$ will still exhibit a
relationship). Thus, the IMPINT estimator tends to be an optimistic
estimator of $\tau(m)$, in the sense that it systematically overstates
predictive accuracy. A simple bias correction addresses this problem.

We observed empirically (see section \ref{lcest}) that for any
positive integer $k$, the estimator $\hat{\tau}_{II}(m) -
\hat{\tau}_{II}(m+k)$ exhibits little bias as an estimator of $\tau(m)
- \tau(m+k)$. That is, the IMPINT estimator is nearly unbiased for the
increments of the learning curve but not necessarily for its overall
level. An explanation for this observation parallels the intuition
behind the bootstrap as follows. The asymptote of the learning curve
$\lim_{m\rightarrow \infty}\tau(m)$ is the Bayes error rate and thus
depends exclusively on $p_{X}(x)$ and $\beta^*$. However, the
increments of $\tau$, in addition to depending on $\beta^*$ and
$p_{X}(x)$, depend on the sampling properties of the estimator
$\hat{\beta}_{m}$ as well as the convergence behavior of
$\hat{\beta}_{m}$ to its limiting value $\beta^*$. The increments of
the IMPINT learning curve estimator $\hat{\tau}_{II}(n)$ are
determined by the sampling properties of $\hat{\beta}_{m}^{(b)}$ as an
estimator of $\bHat$ as well as the manner by which
$\hat{\beta}_{m}^{(b)}$ approaches its limiting value. Thus, if the
sampling properties of $\hat{\beta}_{m}^{(b)}$ about $\bHat$
accurately reflect the sampling properties of $\bHat$ about $\beta^*$
and if $\hat{p}_{X}(x)$ is a reasonable estimator of $p_{X}(x)$, then
it is possible the increments of $\hat{\tau}_{II}$ approximate the
increments of $\tau$.

If one can accurately estimate the increments of the learning curve,
all that remains is to find an unbiased estimator of the learning
curve at a single training set size. This is provided by the
leave-one-out cross-validation (LOOCV) estimator of the expected test
error based on the complete observed training set $\mathcal{D}_n$,
which provides an unbiased estimator of $\tau(n-1)$. Let
$\hat{\tau}_{CV}(n-1)$ denote the LOOCV estimator of $\tau(n-1)$.  The
BRIE is then given by $\hat{\tau}_{B}(m) \triangleq \hat{\tau}_{II}(m)
+ (\hat{\tau}_{CV}(n-1) - \hat{\tau}_{II}(n-1))$.  Thus, BRIE is
simply a shifted version of the IMPINT estimator.

The choice to use the LOOCV estimator of the misclassification rate is
not essential. Noting that $\tau(n) \approx \tau(n-1)$, one could
employ any unbiased (or nearly unbiased) estimator of the expected
test error $\tau(n)$ to recenter the uncorrected estimator of the
learning curve. There are also other ways to achieve this bias
correction. For instance, the MLE $\hat{\beta}_n$ of $\beta$ derived
from the training set may be rescaled by a factor $c<1$, producing a
shrunken coefficient vector $\beta$.  We found that this approach
gives similar results to those obtained using the simple additive bias
correction.

\section{Empirical studies}\label{simstudy}

In this section we examine the performance of the SUBEX, IMPINT, and
BRIE procedures in terms of bias and variance, using a series of
simulation studies.  For each example we used 1000 Monte Carlo
iterations, $B = 1000$ imputed data sets, and an imputed test set
$\mathcal{D}_*$ of size $5000$. All initial training sets are of size
$n=50$ and we estimate $\tau(m)$ for $m = 75, 100, 150,$ and $200$
using this initial training set. Thus, we are attempting to
extrapolate substantially beyond the initial training set size.

For our simulation studies, we consider the following class of models.
The distribution of the features $X$ is a $p$-variate normal
distribution with isotropic variance-covariance matrix $\Sigma$ given
by $\Sigma_{i,j} = r^{|i-j|}$.  The true parameter vector $\beta^*$ is
a $p$-vector of ones.  Thus, this class of models is determined by the
dimension of the model $p$ and the level of dependence between the
features which is governed by the parameter $r\in (-1,1)$.

\begin{figure}[here]
\begin{center}
\begin{minipage}{.45\linewidth}
\includegraphics[width=2.5in]{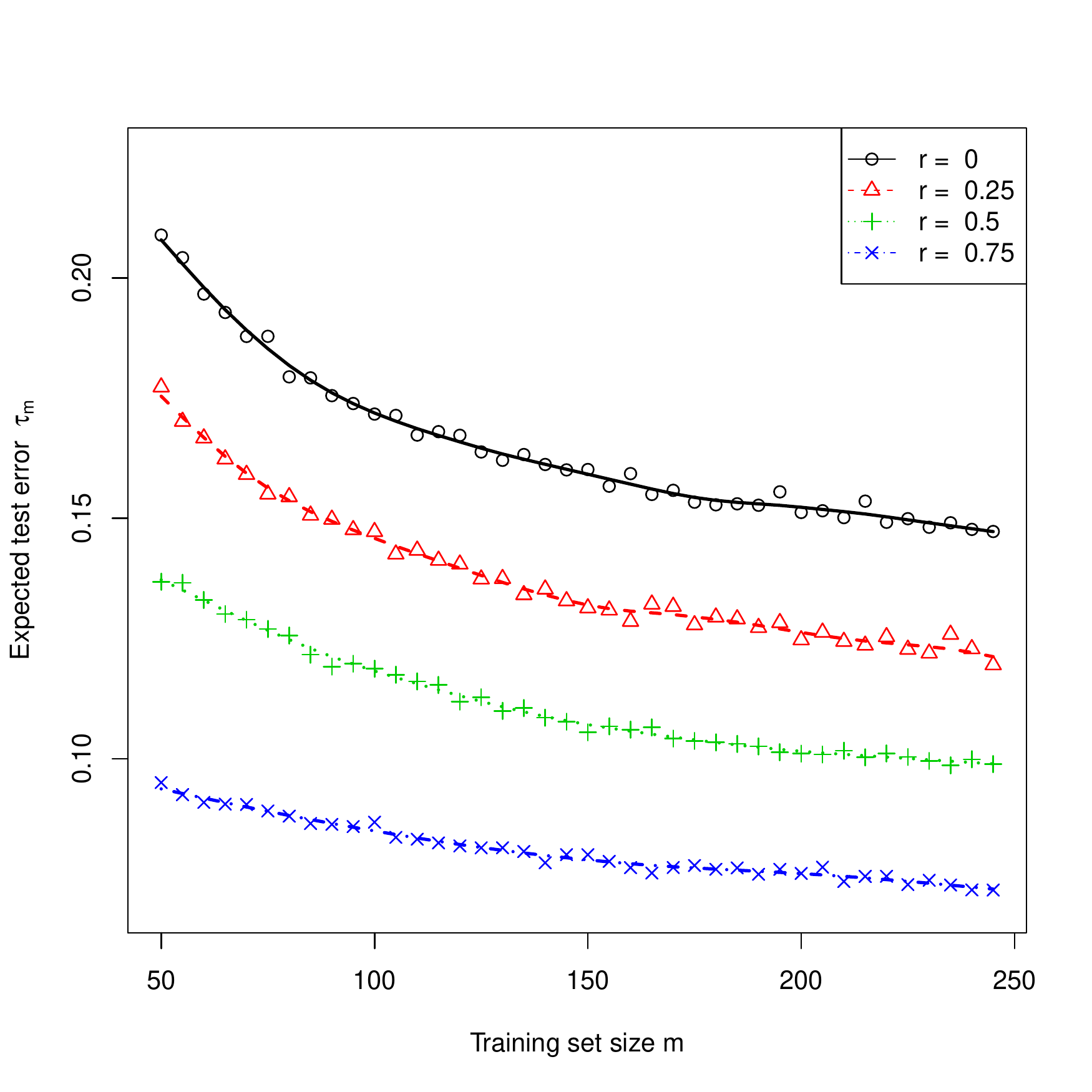}
\end{minipage}
\begin{minipage}{.45\linewidth}
\includegraphics[width=2.5in]{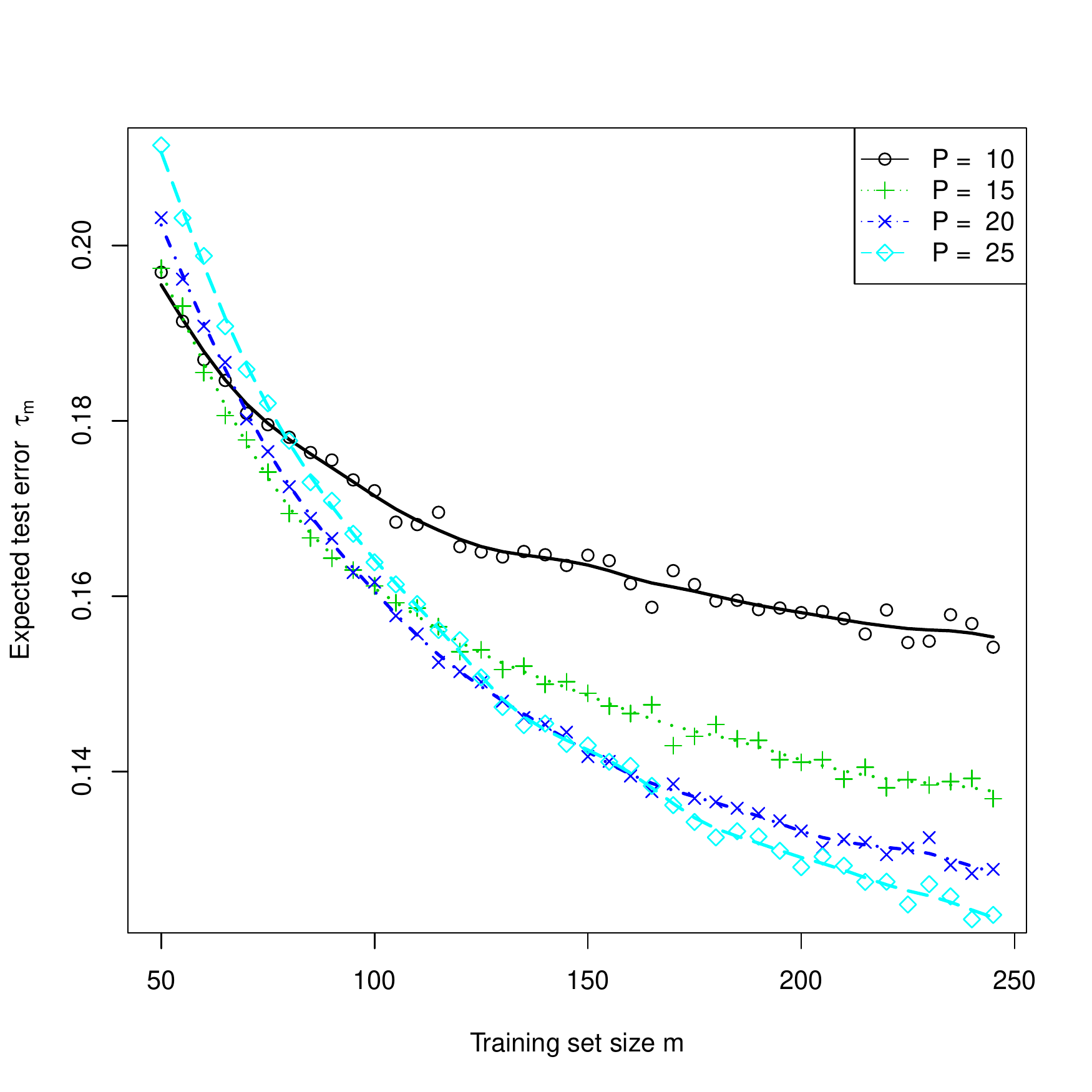}
\end{minipage}
\end{center}
\caption{\textbf{Left:} Learning curves $\tau(m)$ for training sets of
  size $n=50$, for the isotropic normal model with $p=15$ and $r = 0,
  .25, .5, .75$.  \textbf{Right:} Learning curves $\tau(m)$ for
  training sets of size $n=50$ for the isotropic normal model with
  $r=.10$ and $p = 10, 15, 20, 25$.} \label{lcTruths}
\end{figure}

Figure \ref{lcTruths} shows a few examples of learning curves
generated using this class of models.  The left hand side of Figure
\ref{lcTruths} shows the learning curve $\tau(m)$ based on training
sets of size $n=50$, for $p$ fixed at $15$ and $r = 0.0, 0.25, 0.50,$
and $0.75$.  The figure shows that as $r$ increases, both the learning
rate and the asymptotic error rate (i.e.\ the Bayes error rate)
decrease (differences in the learning rate are most evident for sample
sizes less than 100).  These changes are mostly driven by the fact
that as the positive correlation $r$ increases, the distribution of
$|X^{\T}\beta^*|$ becomes stochastically larger, and thus fewer points
lie near the optimal decision boundary $\lbrace x \in \rp\,:\,
x^{\T}\beta^* = 0\rbrace$.  The right hand side of Figure
\ref{lcTruths} shows the learning curve $\tau(m)$ based on training
sets of size $n=50$ for $r$ fixed at $0.10$ and $p = 10, 15, 20$, and
$25$.  The figure shows, as would be expected, that the learning curve
becomes steeper as the dimension of the problem increases (the raw
values of $\tau_m$ are difficult to compare across values of $p$ since
the Bayes error rate changes with $p$).  We use these examples to
examine bias and variance properties of the BRIE and SUBEX estimator.

\subsection{Task one: estimating the improvement in expected error rate}

We first consider the task of estimating the improvement in expected
error rate if additional training data are obtained.  That is, our
goal is to estimate $\delta(n,m) \triangleq \tau(n) - \tau(m)$ for $m
\ge n$.  Note that the plug-in BRIE estimate of $\delta(n, m)$,
$\hat{\delta}_{B}(n,m) \triangleq \hat{\tau}_B(n) - \hat{\tau}_B(n)$,
is equal to the plug-in IMPINT estimate of $\delta(n, m)$.  Estimation
of $\delta(n,m)$ is a somewhat easier problem than estimating the
entire learning curve $\tau(m)$, as it only requires estimating the
shape but not the absolute level of the learning curve.  This quantity
may be of interest to a researcher who cares more about relative
improvement, e.g.\ a 5\% reduction in expected error rate, than
absolute improvement in error rate, e.g.\ a reduction from 12\% to
7\%.

Tables 1 and 2 show the estimated expected values and standard
deviations for the BRIE (IMPINT) and SUBEX estimators, on a class of
eight models as described at the beginning of this section.  The BRIE
exhibits substantially smaller bias and standard deviation than the
SUBEX estimator in all instances, and provides useful estimates even
when extrapolating from $n=50$ to $m=200$, a four-fold increase in
sample size.  The reduction in variability in BRIE relative to SUBEX
presumably results from the variance in SUBEX resulting from the use
of the LOOCV estimator $\hat{\tau}_{CV}(n-1)$, which is well-known to
be highly variable (\cite{toussaint1974}; \cite{efron1983estimating};
\cite{snapinn1985}, \cite{breiman1992}).  The bias in the SUBEX
estimator could be due to the true value of $\tau$ not following an
inverse power law exactly, or could be due to the asymmetric
constraints imposed on the fitted curve, as discussed above.

In tables 1 and 2, the marginal distribution $p_{X}(x)$ was estimated
using maximum likelihood for a normal model with unknown mean vector
$\mu$ and unknown variance-covariance matrix $\Sigma$ given by
$\Sigma_{i,j} = \sigma^2\rho^{|i-j|}$.  Thus, this simulation provides
the BRIE estimator the advantage of knowing the form of the covariance
matrix.

Tables 3 and 4 show the analogous results when $\hat{p}_{X}(x)$ is
modeled as a multivariate normal distribution with unconstrained
variance-covariance matrix. The usual plug-in estimator of the
covariance is used.  The tables show that the BRIE estimator still
exhibits substantially smaller bias and variability than the SUBEX
estimator.  However, the standard deviation of the BRIE is on average
about twice as large compared with using the constrained covariance
estimate as in tables 1 and 2.

\begin{table}
{\small
\begin{center}
\begin{tabular}{cc|c|c|c||c|c|c|c|c|c}
$p$ & $r$ & $\delta(50,75)$ & $\hat{\delta}_{B}(50,75)$, SD &
$\hat{\delta}_{S}(50,75)$, SD & 
$\delta(50,100)$ & $\hat{\delta}_B(50,100)$, SD 
& $\hat{\delta}_{S}(50, 100)$, SD\\ \hline
 $10$ & $0.10$ & $0.0173$ & $0.0145,\,0.00477$ & $0.0254,\, 0.0296$ & 
$0.0249$ & $0.0209,\,0.00613$ & $0.0415,\,0.0486$ \\ 
$15$ & $0.10$ &  $0.0232$ & $0.0253,\,0.00362$ & $0.0297,\,0.0323$ & 
$0.0362$ & $0.0376,\,0.00461$ & $0.0484,\,0.0531$   \\ 
$20$ & $0.10$ & $0.0266$ & $0.0327,\,0.00382$ & $0.0383,\,0.0397$ & 
$0.0415$ & $0.0497,\,0.00499$ & $0.0625,\,0.0650$ \\ 
$25$ & $0.10$ & $0.0294$ & $0.0328,\, 0.00347$ & $0.0400,\,0.0384$ & 
$0.0475$ & $0.0498,\, 0.00450$ & $0.0653,\,0.0630$ \\ 
$15$ & $0.00$ & $0.0210$ & $0.0263,\, 0.00344$ & $0.0296,\,0.0321$ & 
$0.0372$ & $0.0390,\, 0.00440$ & $0.0483,\,0.0527$ \\ 
$15$ & $0.25$ & $0.0222$ & $0.0226,\, 0.00375$ & $0.0272,\,0.0317$ & 
$0.0300$ & $0.0340,\,0.00478$ & $0.0444,\,0.0520$ \\  
$15$ & $0.50$ & $0.0100$ & $0.0169,\, 0.00462$ & $0.0235,\,0.0287$ &
$0.0180$ & $0.0259,\,0.00615$ & $0.0383,\,0.0471$ \\ 
$15$ & $0.75$ & $0.00590$ & $0.00986,\,0.00478$ & $0.0147,\,0.0166$ &
$0.00824$ & $0.0151,\,0.00651$ & $0.0238,\,0.0271$ \\ 
\end{tabular}
\caption{Mean and standard deviation of the BRIE and SUBEX estimator
  for estimating the improvement in the expected error rate when the
  training set size is increased to $n=50$ to $m=75$ and $m=100$.  The
  BRIE estimator used a multivariate normal model with the restriction
  that $\Sigma_{ij} = \sigma^2\rho^{|i-j|}$ to estimate $p_{X}(x)$.
  The BRIE estimator is seen to be significantly less biased than the
  SUBEX estimator, as well as, possessing smaller standard deviation
  across training sets.}
\end{center}
}
\end{table}

\begin{table}
{\small
\begin{center}
\begin{tabular}{cc|c|c|c||c|c|c|c|c|c}
$p$ & $r$ & $\delta(50,150)$ & $\hat{\delta}_{B}(50,150)$, SD &
$\hat{\delta}_{S}(50,150)$, SD & 
$\delta(50,200)$ & $\hat{\delta}_B(50,200)$, SD 
& $\hat{\delta}_{S}(50, 200)$, SD\\ \hline
$10$ & $0.10$ & $0.0322$ & $0.0268,\,0.00679$ & $0.0616,\,0.0727$ & $0.0388$ &
$0.0285,\, 0.00686$ & $0.0742,\,0.0880$ \\ 
$15$ & $0.10$ & $0.0484$ & $0.0497,\, 0.00496$ & $0.0720,\,0.0794$ & $0.0563$ &
$0.0556,\,0.00490$ & $0.0870,\,0.0963$ \\  
$20$ & $0.10$ & $0.0614$ & $0.0673,\, 0.00523$ &  $0.0930,\,0.0971$ & 
$0.0699$ & $0.0765,\, 0.00541$ & $0.112,\,0.117$ \\ 
$25$ & $0.10$ & $0.0684$ & $0.0674,\,0.00484$ & $0.0973,\,0.0943$ & 
$0.0823$ & $0.0767,\, 0.00469$ & $0.117,0.124$ \\ 
$15$ & $0.00$ & $0.0487$ & $0.0512,\, 0.00478$ & $0.0719,\,0.0789$ & 
$0.0576$ & $0.0571,\, 0.00476$ & $0.0868,\,0.0956$ \\  
$15$ & $0.25$ & $0.0459 $ & $0.0456,\,0.00506$ & $0.0660,\,0.0778$ & 
$0.0525$ & $0.0515,\, 0.00487$ & $0.0797,\,0.0943$ \\ 
$15$ & $0.50$ & $0.0313$ & $0.0356,\,0.00684$ & $0.0570,\,0.0705$ & 
$0.0357$ & $0.0410,\,0.00673$ & $0.0687,\,0.0855$ \\ 
$15$ & $0.75$ & $0.0150$ & $0.0210,\,0.00763$ & $0.0351,\,0.0405$ &
$0.0189$ & $0.0244,\,0.00791$ & $0.0421,\,0.0490$ \\  
\end{tabular}
\caption{Mean and standard deviation of the BRIE and SUBEX estimator
  for estimating the improvement in the expected error rate when the
  training set size is increased to $n=50$ to $m=150$ and $m=200$.
  The BRIE estimator used a multivariate normal model with the
  restriction that $\Sigma_{ij} = \sigma^2\rho^{|i-j|}$ to estimate
  $p_{X}(x)$.  The BRIE estimator is seen to be significantly less
  biased than the SUBEX estimator, as well as, possessing smaller
  standard deviation across training sets.}
\end{center}
}
\end{table}

\begin{table}
{\small
\begin{center}
\begin{tabular}{cc|c|c|c||c|c|c|c|c|c}
$p$ & $r$ & $\delta(50,75)$ & $\hat{\delta}_{B}(50,75)$, SD &
$\hat{\delta}_{S}(50,75)$, SD & 
$\delta(50,100)$ & $\hat{\delta}_B(50,100)$, SD 
& $\hat{\delta}_{S}(50, 100)$, SD\\ \hline
 $10$ & $0.10$ & $0.0173$ & $0.0100,\,0.00206$ & $0.0254,\, 0.0296$ & 
$0.0249$ & $0.0140,\,0.00330$ & $0.0415,\,0.0486$ \\ 
$15$ & $0.10$ &  $0.0232$ & $0.0209,\,0.00670$ & $0.0297,\,0.0323$ & 
$0.0362$ & $0.0314,\,0.00865$ & $0.0484,\,0.0531$   \\ 
$20$ & $0.10$ & $0.0266$ & $0.0240,\,0.00372$ & $0.0383,\,0.0397$ & 
$0.0415$ & $0.0372,\,0.00519$ & $0.0625,\,0.0650$ \\ 
$25$ & $0.10$ & $0.0294$ & $0.0256,\, 0.00748$ & $0.0400,\,0.0384$ & 
$0.0475$ & $0.0404,\, 0.0102$ & $0.0653,\,0.0630$ \\ 
$15$ & $0.00$ & $0.0210$ & $0.0217,\, 0.00704$ & $0.0296,\,0.0321$ & 
$0.0372$ & $0.0325,\,0.00910$ & $0.0483,\,0.0527$ \\ 
$15$ & $0.25$ & $0.0222$ & $0.0184,\, 0.00620$ & $0.0272,\,0.0317$ & 
$0.0300$ & $0.0280,\,0.00806$ & $0.0444,\,0.0520$ \\  
$15$ & $0.50$ & $0.0100$ & $0.0140,\, 0.00568$ & $0.0235,\,0.0287$ &
$0.0180$ & $0.0215,\,0.00750$ & $0.0383,\,0.0471$ \\ 
$15$ & $0.75$ & $0.00590$ & $0.00879,\,0.00413$ & $0.0147,\,0.0166$ &
$0.00824$ & $0.0135,\,0.00554$ & $0.0238,\,0.0271$ \\ 
\end{tabular}
\caption{Mean and standard deviation of the BRIE and SUBEX estimator
  for estimating the improvement in the expected error rate when the
  training set size is increased to $n=50$ to $m=75$ and $m=100$.  The
  BRIE estimator used an unrestricted multivariate normal model to
  estimate $p_{X}(x)$.  The BRIE estimator is seen to be significantly
  less biased than the SUBEX estimator, as well as, possessing smaller
  standard deviation across training sets.}
\end{center}
}
\end{table}

\begin{table}
{\small
\begin{center}
\begin{tabular}{cc|c|c|c||c|c|c|c|c|c}
$p$ & $r$ & $\delta(50,150)$ & $\hat{\delta}_{B}(50,150)$, SD &
$\hat{\delta}_{S}(50,150)$, SD & 
$\delta(50,200)$ & $\hat{\delta}_B(50,200)$, SD 
& $\hat{\delta}_{S}(50, 200)$, SD\\ \hline
$10$ & $0.10$ & $0.0322$ & $0.0202,\,0.00476$ & $0.0616,\,0.0727$ & $0.0388$ &
$0.0240,\, 0.00563$ & $0.0742,\,0.0880$ \\ 
$15$ & $0.10$ & $0.0484$ & $0.0421,\, 0.00939$ & $0.0720,\,0.0794$ & $0.0563$ &
$0.0477,\,0.00918$ & $0.0870,\,0.0963$ \\  
$20$ & $0.10$ & $0.0614$ & $0.0519,\, 0.0101$ &  $0.0930,\,0.0971$ & 
$0.0699$ & $0.0603,\, 0.00998$ & $0.112,\,0.117$ \\ 
$25$ & $0.10$ & $0.0684$ & $0.0578,\,0.0118$ & $0.0973,\,0.0943$ & 
$0.0823$ & $0.0682,\, 0.0118$ & $0.117,0.124$ \\ 
$15$ & $0.00$ & $0.0487$ & $0.0435,\, 0.00991$ & $0.0719,\,0.0789$ & 
$0.0576$ & $0.0492,\,0.00974$ & $0.0868,\,0.0956$ \\  
$15$ & $0.25$ & $0.0459 $ & $0.0381,\,0.00878$ & $0.0660,\,0.0778$ & 
$0.0525$ & $0.0436,\, 0.00857$ & $0.0797,\,0.0943$ \\ 
$15$ & $0.50$ & $0.313$ & $0.0298,\,0.00839$ & $0.0570,\,0.0705$ & 
$0.0357$ & $0.0345,\,0.00836$ & $0.0687,\,0.0855$ \\ 
$15$ & $0.75$ & $0.0150$ & $0.0188,\,0.00636$ & $0.0351,\,0.0405$ &
$0.0189$ & $0.0218,\,0.00650$ & $0.0421,\,0.0490$ \\  
\end{tabular}
\caption{Mean and standard deviation of the BRIE and SUBEX estimator
  for estimating the improvement in the expected error rate when the
  training set size is increased to $n=50$ to $m=150$ and $m=200$.
  The BRIE estimator used an unrestricted multivariate normal model to
  estimate $p_{X}(x)$.  The BRIE estimator is seen to be significantly
  less biased than the SUBEX estimator, as well as, possessing smaller
  standard deviation across training sets.}
\end{center}
}
\end{table}

\subsection{Task two: estimating the learning curve}

Next we consider the more difficult task of estimating the full
learning curve, not just its increments.  Our goal is to estimate
$\tau(75), \tau(100), \tau(150),$ and $\tau(200)$ using a training set
size of $n=50$.  As in the previous section, we use the SUBEX
estimator as a baseline for comparison.

Tables (\ref{LCSmallSamplesII}) and (\ref{LCLargeSamplesII}) show the
estimated expected values and standard deviations for the BRIE and the
SUBEX estimator of the learning curve on the same eight models
considered in the preceding section.  Like the results for estimating
the improvement in error rate, the BRIE shows only negligible bias for
estimating the learning curve, while the bias for SUBEX is
substantial.  BRIE also has a smaller variance than SUBEX, but the
advantage is substantially smaller than in the case of estimating
learning curve increments.  This is likely due to the fact that the
bias correction used in the BRIE method is based on the highly
variable leave-one-out cross validation estimator of $\tau(n-1)$.  

In practice, learning curve estimates are useful to the extent that
they can distinguish between substantially different possible true
learning curve patterns.  For estimating learning curve increments,
with an increase in sample size from 50 to 150 (table 4, left
columns), the range of possible true increments is roughly 0.05 (0.015
to 0.0684).  The standard error for the BRIE estimate of these
quantities is around 0.01, so the range of possible outcomes in our
simulations is at least 5 times greater than the standard error.  For
estimating the learning curves themselves, the analogous results in
table 6 (left columns) show a range of 0.08 in the true values, and a
standard error of $0.05-0.07$.  Thus the maximum observed difference is
only slightly greater than the standard error, suggesting that the
practical value of estimators of the learning curve may be limited,
while useful information can be obtained from estimates of the
learning curve increments.  This point is underscored in the example
considered in the next section.


\begin{table}
{\small
\begin{center}
\begin{tabular}{cc|c|c|c||c|c|c|c|c|c}
$p$ & $r$ & $\tau(75)$ & $\hat{\tau}_{B}(75)$, SD &
$\hat{\tau}_{S}(75)$, SD &
$\tau(100)$ & $\hat{\tau}_B(100)$, SD
& $\hat{\tau}_{S}(100)$, SD\\ \hline
 $10$ & $0.10$ & $0.179$ & $0.184,\,0.0634$ & $0.166,\, 0.0735$ &
$0.172$ & $0.179,\,0.0619$ & $0.152,\,0.0828$ \\ 
$15$ & $0.10$ &  $0.174$ & $0.175,\,0.0636$ & $0.141,\,0.0755$ &
$0.161$ & $0.164,\,0.0639$ & $0.127,\,0.0832$   \\ 
$20$ & $0.10$ & $0.176$ & $0.177,\,0.0679$ & $0.154,\,0.0831$ &
$0.161$ & $0.163,\,0.0678$ & $0.136,\,0.0926$ \\ 
$25$ & $0.10$ & $0.181$ & $0.187,\, 0.0715$ & $0.174,\,0.0851$ &
$0.163$ & $0.172,\, 0.0713$ & $0.151,\,0.0994$ \\ 
$15$ & $0.00$ & $0.187$ & $0.190,\, 0.0655$ & $0.169,\,0.0831$ &
$0.171$ & $0.179,\, 0.0664$ & $0.151,\,0.0940$ \\ 
$15$ & $0.25$ & $0.155$ & $0.158,\, 0.0610$ & $0.147,\,0.0739$ &
$0.147$ & $0.148,\,0.0610$ & $0.134,\,0.0814$ \\  
$15$ & $0.50$ & $0.126$ & $0.126,\, 0.0581$ & $0.110,\,0.0686$ &
$0.118$ & $0.118,\,0.0582$ & $0.100,\,0.0748$ \\ 
$15$ & $0.75$ & $0.0891$ & $0.0861,\,0.0507$ & $0.0744,\,0.0505$ &
$0.0868$ & $0.0814,\,0.0508$ & $0.0683,\,0.0527$ \\ 
\end{tabular}
\caption{Mean and standard deviation of the BRIE and SUBEX estimator
  for estimating the learning curve when the training set size is
  increased from $n=50$ to $m=75$ and $m=100$.  The BRIE estimator
  used a multivariate normal model with the restriction that
  $\Sigma_{ij} = \sigma^2\rho^{|i-j|}$ to estimate $p_{X}(x)$.  The
  BRIE estimator is seen to be significantly less biased than the
  SUBEX estimator, as well as, possessing smaller standard deviation
  across training sets.}\label{LCSmallSamplesII}
\end{center}
}
\end{table}

\begin{table}
{\small
\begin{center}
\begin{tabular}{cc|c|c|c||c|c|c|c|c|c}
$p$ & $r$ & $\tau(150)$ & $\hat{\tau}_{B}(150)$, SD &
$\hat{\tau}_{S}(150)$, SD &
$\tau(200)$ & $\hat{\tau}_B(200)$, SD
& $\hat{\tau}_{S}(200)$, SD\\ \hline
 $10$ & $0.10$ & $0.164$ & $0.173,\,0.0647$ & $0.139,\, 0.0911$ &
$0.158$ & $0.169,\,0.0651$ & $0.133,\,0.0945$ \\ 
$15$ & $0.10$ &  $0.149$ & $0.153,\,0.0643$ & $0.113,\,0.0905$ &
$0.141$ & $0.146,\,0.0644$ & $0.107,\,0.0930$   \\ 
$20$ & $0.10$ & $0.141$ & $0.148,\,0.0676$ & $0.120,\,0.100$ &
$0.133$ & $0.139,\,0.0674$ & $0.112,\,0.103$ \\ 
$25$ & $0.10$ & $0.143$ & $0.155,\, 0.0710$ & $0.130,\,0.111$ &
$0.129$ & $0.144,\, 0.0707$ & $0.122,\,0.114$ \\ 
$15$ & $0.00$ & $0.160$ & $0.167,\, 0.0670$ & $0.134,\,0.101$ &
$0.151$ & $0.161,\, 0.0674$ & $0.127,\,0.103$ \\ 
$15$ & $0.25$ & $0.131$ & $0.137,\, 0.0611$ & $0.121,\,0.116$ &
$0.124$ & $0.131,\,0.0613$ & $0.116,\,0.0917$ \\  
$15$ & $0.50$ & $0.106$ & $0.101,\, 0.0583$ & $0.919,\,0.0795$ &
$0.101$ & $0.105,\,0.0583$ & $0.0886,\,0.0811$ \\ 
$15$ & $0.75$ & $0.0800$ & $0.0761,\,0.0508$ & $0.0625,\,0.0550$ &
$0.0761$ & $0.0730,\,0.0508$ & $0.0597,\,0.0561$ \\ 
\end{tabular}
\caption{Mean and standard deviation of the BRIE and SUBEX estimator
  for estimating the learning curve when the training set size is
  increased from $n=50$ to $m=150$ and $m=200$.  The BRIE estimator
  used a multivariate normal model with the restriction that
  $\Sigma_{ij} = \sigma^2\rho^{|i-j|}$ to estimate $p_{X}(x)$.  The
  BRIE estimator is seen to be significantly less biased than the
  SUBEX estimator, as well as, possessing smaller standard deviation
  across training sets.}\label{LCLargeSamplesII}
\end{center}
}
\end{table}

\section{Example: Predicting the four year survival probability for CLL}\label{cll}

We next demonstrate the BRIE approach to estimating learning curves
using data from a study of prognostic factors for outcomes of patients
with chronic lymphocytic leukemia (CLL).  The duration between the
time that a subject entered a study and the time the subject required
treatment (TTFT), a surrogate for disease progression, was obtained
for 209 CLL subjects in a prospective study \cite{sheddenmalek}.  For
this analysis, the TTFT outcomes were dichotomized according to
whether treatment was needed within four years of diagnosis.  Eleven
potential prognostic markers were used to predict this outcome using
logistic regression.  The markers are: ZAP70\%, p53 status, CD38\%,
IgVH mutation status, age at diagnosis, Rai stage at diagnosis, number
of positive lymph node groups, subchromosomal losses, chromosomal
losses, subchromosomal gains, and chromosomal gains.  This set of
predictor variables includes binary (p53, IgVH), ordered categorical
(Rai stage), count (chromosomal/subchromosomal losses and gains,
positive lymph node groups), and continuous (ZAP70\%, CD38\%, age)
measures.

We calculated the BRIE estimator of the learning curves using two
different models for the marginal distribution $p_X(x)$ of predictor
variables.  The first approach, which we denote ``GM'', used a
Gaussian mixture.  Here we stratified the training data into four
groups, according to the joint pattern of values for the two binary
variables (p53 and IgVH).  Within each such group, the mean of the
other nine non-binary variables was calculated, and a pooled
covariance matrix for these nine variables over the four groups
(centered at their respective means) was calculated.  The GM model for
the predictor variables was a four component Gaussian mixture, where
the four components correspond to the four strata determined by the
joint values of p53 and IgVH.  The mixture components had means equal
to the four stratum means in the training data, a common covariance
structure equal to the pooled covariance matrix from the data, and
marginal frequencies equal to the empirical frequencies of the four
subgroups of training data defined by p53 and IgVH.

The second approach to modeling the covariate distribution, which we
denote ``GC'', used a Gaussian copula.  Here all data were converted
to normal scores, then the correlation matrix of the normal scores was
calculated.  To produce simulated data from this model, we simulated
Gaussian vectors according to this covariance matrix, then transformed
each component of these vectors with the corresponding inverse normal
scores function.  The resulting model for simulated data has marginal
distributions exactly equal to the univariate empirical distributions
of the training set, and dependence which approximates the training
set dependence.

The results are given in table \ref{table:cll}.  Four different
training set sizes are used (50, 75, 100, 150).  For each training set
size $n$, we sampled $n$ observations without replacement from the
overall CLL data set of 209 observations. These $n$ observations were
used for three purposes: to estimate the misclassification error rate
using cross-validated logistic regression, to fit a logistic
regression model for $p_{Y|X}(y|x)$, and to fit models for the
predictor variable distribution $p_X(x)$ using the GM and GC
approaches.  Together, $p_X(x)$ and $p_{Y|X}(y|x)$ were used to define
the data-generating population $p^*_{Y,X}(y,x) = p_{Y|X}(y|x)p_X(x)$.
We then generated data sets of size $n$, $2n$, and $3n$ from
$p^*_{Y,X}(y,x)$, fit another logistic regression model to each of
these data sets, and evaluated the accuracy of these fitted rules
relative to the data-generating population $p^*_{Y,X}(y,x)$.  This
process was repeated 1000 times and averaged to produce the results in
table \ref{table:cll}.

Table \ref{table:cll} shows that improvement in prediction accuracy of
only 2-3\% may occur when increasing the training set sample size by
factors of 2-3 in this setting.  This result is stable over the two
approaches to modeling the predictor variables.  As expected, the
magnitudes of the gain are greatest for smaller training set sizes,
and when the increment in training set size is larger (i.e.\ when
comparing $\hat{\tau}_{II}(3n)-\hat{\tau}_{II}(n)$ to
$\hat{\tau}_{II}(2n) - \hat{\tau}_{II}(n)$).  Overall, this analysis
suggests that only a small improvement in accuracy is likely to result
from increasing the training set size in this setting.  To achieve
more substantial gains in accuracy, more informative markers, or a
better modeling framework for these 11 markers should be sought.

As expected, $\hat{\tau}_{II}$ underestimates the error rate,
especially when the true error rate is high.  This is a consequence of
overfitting, since the strength of association in the fitted logistic
regression model ${\rm logit} P(Y=1|X=x) = \hat{\beta}^\prime x$ will
tend to be stronger that the strength of association in the true model.
In particular, if $Y$ and $X$ are independent in the true model, the
fitted model will still have some association.  As expected, this
tendency diminishes as the sample size grows.  As a result,
$\hat{\tau}_{II}(m)$ tends to increase with $m$, whereas the CV error,
and presumably the true error, decrease with $n$.  We note that for
larger sample sizes, the copula model for $p_X(x)$ produces BRIE curve
estimates that more closely resemble the cross-validation results.

As noted above, $\hat{\tau}_{II}$ is generally too variable to be
useful, so we focus on the accuracy gain, as shown in the final two
columns of table \ref{table:cll}.  Since we have 209 data points to
work with, we directly apply cross-validation on subsamples up to size
209 to provide a direct cross-validation based estimate that is known
to be nearly unbiased.  For example, under the GC model, the error
rate is predicted to drop from $0.2181$ to $0.1828$ when the training
set sample size grows from 50 to 150, a gain of $0.0353$.  According
to the cross-validation estimates, the gain is $0.2775 - 0.2363 =
0.0412$.  Similarly, when the training set sample size grows from 75
to 150, the predicted gain in classification accuracy using the GC
model for $p_X(x)$ is $0.0191$, and the corresponding estimate from
cross-validation is $0.0273$.

\begin{table}
\begin{center}
\begin{tabular}{rrrrrrrr}
&&&\multicolumn{3}{c}{$\hat{\tau}_{II}$}\\\cline{4-6}
& \multicolumn{1}{c}{$n$} & \multicolumn{1}{c}{CV} &
\multicolumn{1}{c}{$n$} & \multicolumn{1}{c}{$2n$} & 
\multicolumn{1}{c}{$3n$}\\\hline 
GM & 50 & 0.2775 & 0.2210 & 0.1968 & 0.1881 & 0.0242 & 0.0329\\
GM & 75 & 0.2636 & 0.2342 & 0.2157 & 0.2096 & 0.0185 & 0.0246\\
GM & 100 & 0.2580 & 0.2340 & 0.2201 & 0.2150 & 0.0139 & 0.0190\\
GM & 150 & 0.2363 & 0.2364 & 0.2264 & 0.2227 & 0.0100 & 0.0137\\
GC & 50 & 0.2775 & 0.2181 & 0.1922 & 0.1828 & 0.0259 & 0.0353\\
GC & 75 & 0.2636 & 0.2387 & 0.2196 & 0.2130 & 0.0191 & 0.0257\\
GC & 100 & 0.2580 & 0.2467 & 0.2316 & 0.2263 & 0.0151 & 0.0204\\
GC & 150 & 0.2363 & 0.2470 & 0.2358 & 0.2321 & 0.0112 & 0.0149\\\hline
\end{tabular}
\caption{Learning curve analysis for the CLL data. Columns 7 and 8
  show the gain in accuracy when the training set sample size
  increases from $n$ to $2n$ (i.e.\ $\hat{\tau}_{II}(2n) -
  \hat{\tau}_{II}(n)$), and from $n$ to $3n$
  (i.e.\ $\hat{\tau}_{II}(3n)-\hat{\tau}_{II}(n)$),
  respectively. \label{table:cll}}
\end{center}
\end{table}

\vspace{2cm}

\section{Discussion}\label{discussion}

We have discussed three relatively simple approaches for estimating
the learning curve of a classifier.  SUBEX methods rely on a
parametric model of the learning curve, and use unbiased estimates of
the classification error rate for sample sizes smaller than the
observed training set size to estimate the model parameters.  IMPINT
methods model the data distribution, from which the learning curve can
be estimated at arbitrary sample sizes without the need to model the
learning curve.

Learning curve estimation is a challenging problem, and neither method
considered here gives highly accurate results.  However, we see that
even in the limited range of settings considered here, gains in
predictive performance ranging from 0.02 to 0.07 can be observed for
three-fold increases in the training set size (table 4, column 2).
For a problem where predictive accuracies in the 0.8-0.95 range are
typical, knowing that a gain of 0.07 can be achieved may lead to a
very different strategy for follow-up research compared to knowing
that only a gain of 0.02 should be expected. The BRIE approach
estimates the gain in predictive performance nearly unbiasedly, with a
standard error of at most 0.01.  This provides us with power to
confidently assess whether we are at the low end or the high end of
the range of possible gains in performance.

The SUBEX and IMPINT approaches differ in several major ways, any of
which could impact their performances.  One potential drawback of the
SUBEX approach is that the inverse power law model for $\tau$ may not
be able to represent the true learning curve.  An exact analytic
expression for $\tau$ is unlikely to exist, necessitating the use of
convenience parameterizations such as the inverse power law.  Another
concern for SUBEX is its use of cross-validation, which is known to
have high variance (\cite{efron1983estimating}).  This variance may
propagate to the learning curve estimate.  The IMPINT method is not
subject to these limitations, but concerns may arise about the need to
estimate the data generating model, which is not necessary for the
SUBEX approach.  It is unclear which features of the data generating
model are critical for learning curve estimation.  At a minimum, the
dimension and some measure of the strength of the predictive
relationship are clearly relevant.

As noted above, the SUBEX approach models the learning curve, while
the IMPINT approach models the full data distribution.  The learning
curve is a simpler object than the data distribution, hence SUBEX
seems to require fewer assumptions.  However, the learning curve is
not directly observed.  Any appropriate statistical modeling framework
can be used to attain an estimate of the data generating model, and
diagnostic and other tools are available to assess the fit of the
model.  Analogous tools for assessing the appropriateness of the
learning curve model used by the SUBEX procedure are not readily
available.  Furthermore, the errors in the SUBEX procedure will be
amplified by the need to extrapolate beyond the range of sample sizes
that are directly estimated using subsampling.  No analogous source of
variation seems to be present in the IMPINT approach.

Learning curves have the potential to become a useful tool in applied
statistics.  One relevant analogy is to the widely-practiced fields of
power analysis and sample size planning.  In this setting, preliminary
estimates of effect sizes are used to assess the power for various
study designs.  Learning curves can be viewed as a power analysis tool
to be used when the research aims involve prediction, rather than
focusing on estimation and hypothesis testing.  As in classical power
analysis, over-reliance on point estimates from small pilot studies
may not be advised.  In practice it would be advisable to consider a
range of possibilities for key population parameters and attempt to
delineate those situations where substantial gains in predictive
performance are expected to occur.

\bibliographystyle{plainnat}
\bibliography{lc2}
\nocite{mukherjee2003estimating}
\nocite{efron1983estimating}
\nocite{trevor2009elements}
\nocite{insel2009translating}
\nocite{sheddenmalek}
\end{document}